\documentclass[letterpaper, 10 pt, conference]{ieeeconf}  %

\IEEEoverridecommandlockouts                              %
\usepackage{graphicx, gensymb} 
\usepackage{subfigure}
\usepackage{amsmath}
\usepackage{epsfig} 
\usepackage{mathptmx} 
\usepackage{times} 
\usepackage{amssymb}  
\usepackage{balance}
\usepackage{comment}
\usepackage{booktabs}                     
\usepackage{multirow,rotating, array,booktabs}
\usepackage{tikz}
\usetikzlibrary{intersections}
\usetikzlibrary{arrows,shapes,positioning}
\usetikzlibrary{decorations.markings}
\tikzset{->-/.style={decoration={
  markings,
  mark=at position .5 with {\arrow{>}}},postaction={decorate}}}
\title{\LARGE \bf Heterogeneity in Neuronal Calcium Spike Trains based on Empirical Distance}

\author{Sathish Ande$^{*1}$, Jayanth R Regatti$^1$, Neha Pandey$^1$, Ajith Karunarathne$^2$, Lopamudra Giri$^1$,  Soumya Jana$^1$\\${}^1$Indian Institute of Technology Hyderabad, Telangana, India $^2$The University of Toledo, Ohio, USA \\$^*$Email: ee15resch02003@iith.ac.in}

\begin{document}

\maketitle
\thispagestyle{empty}
\pagestyle{empty}

\begin{abstract}
Statistical similarities between neuronal spike trains could reveal significant information on complex underlying processing. In general, the similarity between synchronous spike trains is somewhat easy to identify. However, the similar patterns also potentially appear in an asynchronous manner. However, existing methods for their identification tend to converge slowly, and cannot be applied to short sequences. In response, we propose Hellinger distance measure based on empirical probabilities, which we show to be as accurate as existing techniques, yet faster to converge for synthetic as well as experimental spike trains. Further, we cluster pairs of neuronal spike trains based on statistical similarities and found two non-overlapping classes, which could indicate functional similarities in neurons. Significantly, our technique detected functional heterogeneity in pairs of neuronal responses with the same performance as existing techniques, while exhibiting faster convergence. We expect the proposed method to facilitate large-scale studies of functional clustering, especially involving short sequences, which would in turn identify signatures of various diseases in terms of clustering patterns.

\end{abstract}

\section{Introduction}
 Neurons encode stimulus information in spike trains. In fact, heterogeneity in spike trains is a  known manifestation of complex information processing, which enables diverse functions in the hippocampus, a brain region associated with memory and learning \cite{soltesz2018ca1}. The said heterogeneity in spike trains has been investigated by clustering neuron pairs based on certain statistical similarities. An early attempt in this direction was based on a correlation-based similarity measure \cite{schreiber2003new}. However, such a measure captures coincident firing, i.e., synchronicity in spike trains, but ignores time-delayed versions of similar patterns which are known to arise in complex neuronal networks. As a remedy, distance measures based on Lempel-Ziv (LZ) encoding have been suggested to identify the statistical similarities in synchronous or asynchronous spike trains \cite{christen2004spike,christen2006measuring}. One such method was based on LZ-78 algorithm which needs long sequences for reliable performance. In the quest for a method that can be applied to short sequences, we consider LZ-76, a LZ-based fast method, but find it to be inaccurate. Against this backdrop, we propose a Hellinger distance measure based on empirical probabilities of patterns in each pair of spike trains \cite{sathish2020}. Our method converges faster than LZ-78, and hence may be used on short sequences, while being comparably accurate. Further, we cluster pairs of neuronal spike trains and found two non-overlapping classes, and the clusters obtained using the proposed distance measure and the distance based on LZ-78 are found to behave similarly. This demonstrates the suitability of the proposed method as a fast-converging alternative to the existing slow technique.

 The rest of this paper is organized as follows. Section~\ref{sec:materials} describes calcium imaging of hippocampal neurons and spike train inference,  and introduces the notion of LZ distance and the proposed empirical distance measure. Further, Section~\ref{sec:results} presents the results demonstrating suitability of the proposed method. Finally, Section~\ref{sec:conclusion} concludes the paper.
 
\begin{figure}[t!]
\centering
\includegraphics[width=0.4\columnwidth,height=0.1\textheight]{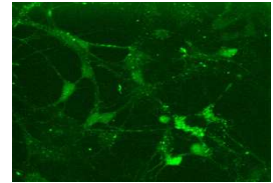}
\label{fig:calciumImaging}
\vspace{-0.9em}
\caption{Intracellular calcium imaging: Representative image of hippocampal neuron population with 28 neurons. Scale bar = 20 $\mu$m \cite{sathish2020}.}
\label{fig:Calcium Imaging}
\end{figure}

\section{Materials and Methods}
\label{sec:materials}
\usetikzlibrary{shapes.geometric, arrows}
\tikzstyle{startstop} = [rectangle, rounded corners, minimum width=3cm, minimum height=0.8cm,text centered, draw=black]
\tikzstyle{io} = [trapezium, trapezium left angle=70, trapezium right angle=110, minimum width=3cm, minimum height=1.5cm, text centered, draw=black, fill=blue!30]
\tikzstyle{process} = [rectangle, minimum width=3cm, minimum height=1cm, text centered, draw=black, fill=orange!30]
\tikzstyle{decision} = [diamond, minimum width=3cm, minimum height=1cm, text centered, draw=black, fill=green!30]
\tikzstyle{arrow} = [thick,->,>=stealth]

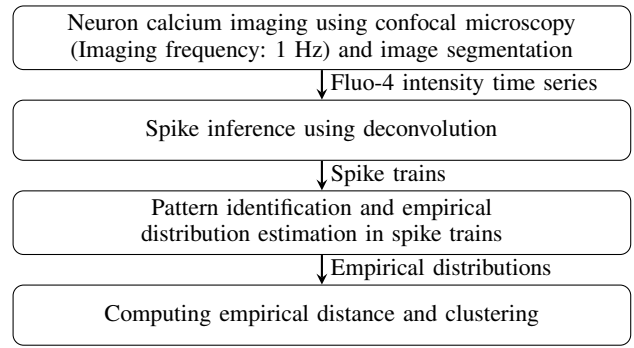
\begin{figure}[t!]
    \centering
   \small
 \begin{tikzpicture}[node distance=35pt]
\node (1) [startstop,text width = 8cm] {Neuron calcium imaging using confocal microscopy (Imaging frequency: 1 Hz) and image segmentation};
\node (2) [startstop,below of =1,text width = 8cm] {Spike inference using deconvolution};
\node (3) [startstop,below of =2,text width = 8cm] {Pattern identification and  empirical distribution estimation in spike trains};
\node (4) [startstop,below of =3,text width = 8cm] {Computing empirical distance and clustering};
\draw [arrow] (1) --node [anchor=west] { Fluo-4 intensity time series} (2) ;
\draw [arrow] (2) --node [anchor=west] {Spike trains}(3);
\draw [arrow] (3) -- node [anchor=west] {Empirical distributions}(4);
\end{tikzpicture}
    \caption{Schematic  workflow.}
    \label{fig:workflow}
\end{figure}
The workflow of the paper is schematically depicted in Fig. \ref{fig:workflow}, and elaborated in the following.

\subsection{Data Collection and Spike Train Inference}
\label{ssec:calcium Imaging}

\begin{figure}[t!]
\centering
\includegraphics[width=\columnwidth,height=0.17\textheight]{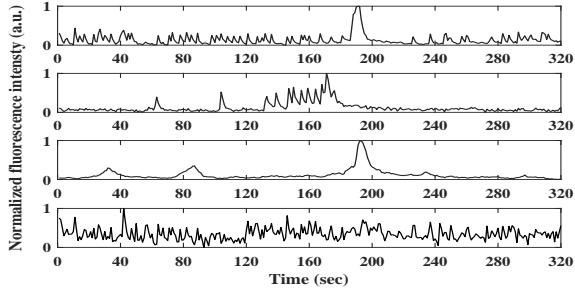}
 \vspace{-2em}
\caption{Time course of normalized Fluo-4 intensity for four neurons indexed 1-4 in a neuron population.}
\label{fig:fluorescence-2}
\end{figure}

We performed time-lapse confocal imaging (using a Leica DMI6000B inverted microscope fitted with a Yokogawa CSU-X1 spinning-disk unit)
on hippocampal neurons, cultured from l day postnatal Sprague-Dawley rats. In particular, we monitored intracellular calcium at $7$-th day after plating using excitation at 488 nm and emission at 510 nm \cite{giri2014g}. During imaging, neurons were kept in the attached incubation chamber maintained at 37$^o$C and 5\% CO$_2$. The interval between successive images, while set at 1 s, was observed to vary between 0.8 s to 1 s. From the time-lapse image data (see Fig. \ref{fig:Calcium Imaging} for a representative frame), the time course of spatially resolved Fluo-4 fluorescence intensity in neuron populations was obtained using Andor software. At present, out of a population of 28 neurons, we consider 8 neurons, indexed 1--8, for analysis (time course of calcium responses are shown in Fig. \ref{fig:fluorescence-2} for neurons 1--4 and heterogeneity in such responses is visually evident here). For each neuron, we inferred binary spike train from its time course using suitable normalization and a fast nonnegative deconvolution algorithm \cite{sathish2020,vogelstein2010fast}. Such spike sequences were used for further analysis.

\subsection{Existing LZ-based Dissimilarity Measures}
Versions (LZ-78, LZ-77, LZ-76) of LZ encoding scheme are based on suitable dictionaries that convert a given sequence $X^n=(X_1,X_2\hdots,X_n)$ into non-overlapping phrases \cite{sathish2020,amigo2004estimating}. For instance, $X^n=`0011001010100111$' is parsed as \\
\indent {LZ-78}: 0|01|1|00|10|101|001|11  \\
\indent {LZ-77}: 0|01|1|10|0010|010|101|010| \\
\indent \indent 101|0100|10011|00111|0111|111|11|1 \\
\indent {LZ-76}: 0|01|10|010|10100|111. \\
The complexity of a sequence $X^n$
is defined as
\begin{equation}
K(X^{n})= \frac{c(X^{n})\log(c(X^{n}))}{n},
\label{eq:lzc}
\end{equation}

 where $c(X^{n})$ is the number of phrases in the dictionary. For each version (LZ-78, LZ-77, LZ-76), the dictionary is different, and hence the complexity defined by (\ref{eq:lzc}) is different. However, each version of complexity
$K(X^{n})$ is known to approach the entropy rate $ \frac{1}{n}H(X^n)$ as $n\rightarrow \infty$, albeit at with a slow rate of convergence \cite{sathish2020}.
 
 For two bit strings $X^n$ and $Y^n$ of equal length, the generic Lempel-Ziv distance is defined as \cite{christen2004spike}
\begin{equation}
d_{\mbox{\scriptsize LZ}} = 
1-\min \Bigg({\frac{K(X^n)-K(X^n|Y^n)}{K(X^n)},\frac{K(Y^n)-K(Y^n|X^n)}{K(Y^n)}}\Bigg),
\label{eq:dLZ}
\end{equation}
where $X^n|Y^n$ contains the phrases in $X^n$ that are not in $Y^n$. Specializing respectively to the versions LZ-78, LZ-77, LZ-76, we define by (\ref{eq:dLZ}) distances $d_{\mbox{\scriptsize LZ78}}$, $d_{\mbox{\scriptsize LZ77}}$, $d_{\mbox{\scriptsize LZ76}}$, considering the corresponding suitable dictionaries.
\begin{figure}[t!]
\centering
\includegraphics[width=\columnwidth]{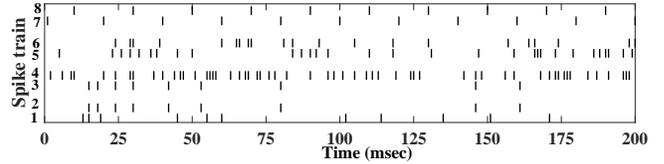}
 \vspace{-2em}
\caption{Raster plots of synthetic spike trains for sequence length 200 (we use length of 500 for analysis) for synthetic trains 1-2 of pair (i), 3-4 of pair (ii), 5-6 of pair (iii) and 7-8 of pair (iv).  }
\label{fig:RPsynthetic}
\end{figure}
 \begin{figure*}[t!]
\centering
\begin{tabular}{cc}
\includegraphics[width=0.9\textwidth]{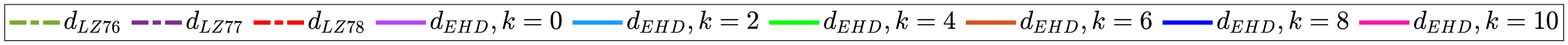}\\
\vspace{-1.8em}\\ 
\subfigure[]{
\includegraphics[width=4.2cm]{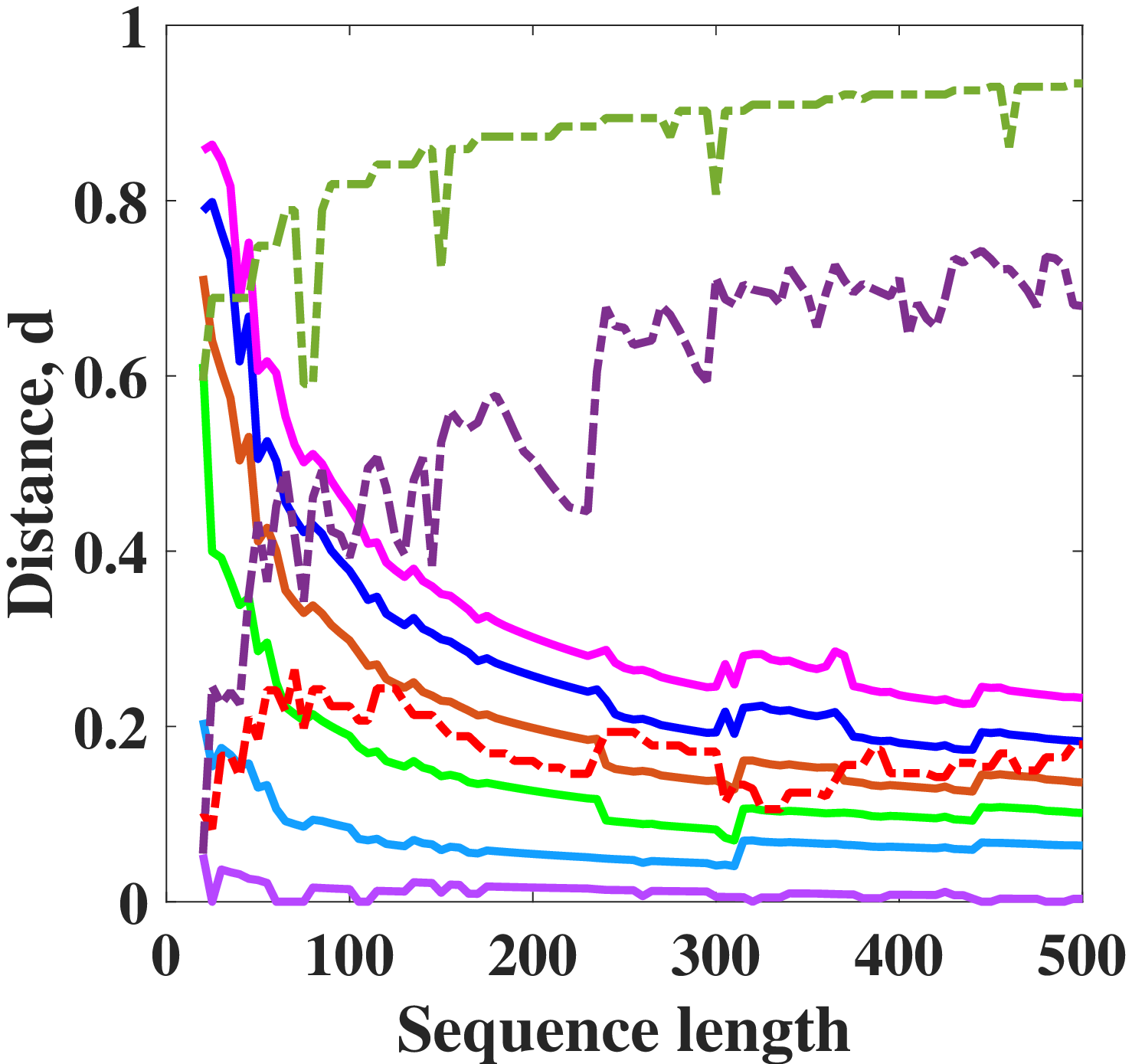}
\label{fig:nd1}\hspace{-1em}}
\subfigure[]{
\includegraphics[width=4.2cm]{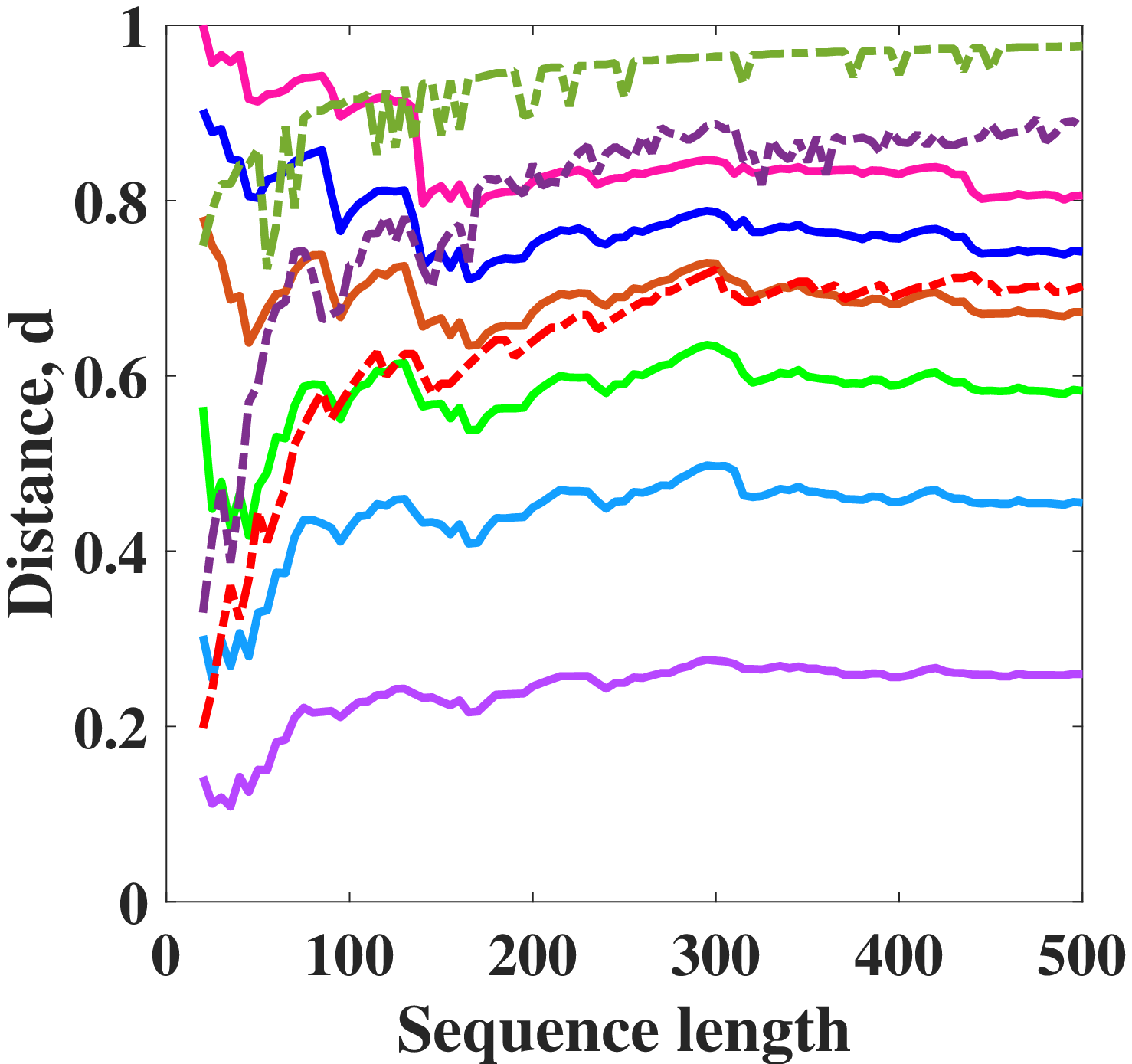}
\label{fig:nd2}\hspace{-1em}} 
\subfigure[]{
\includegraphics[width=4.2cm]{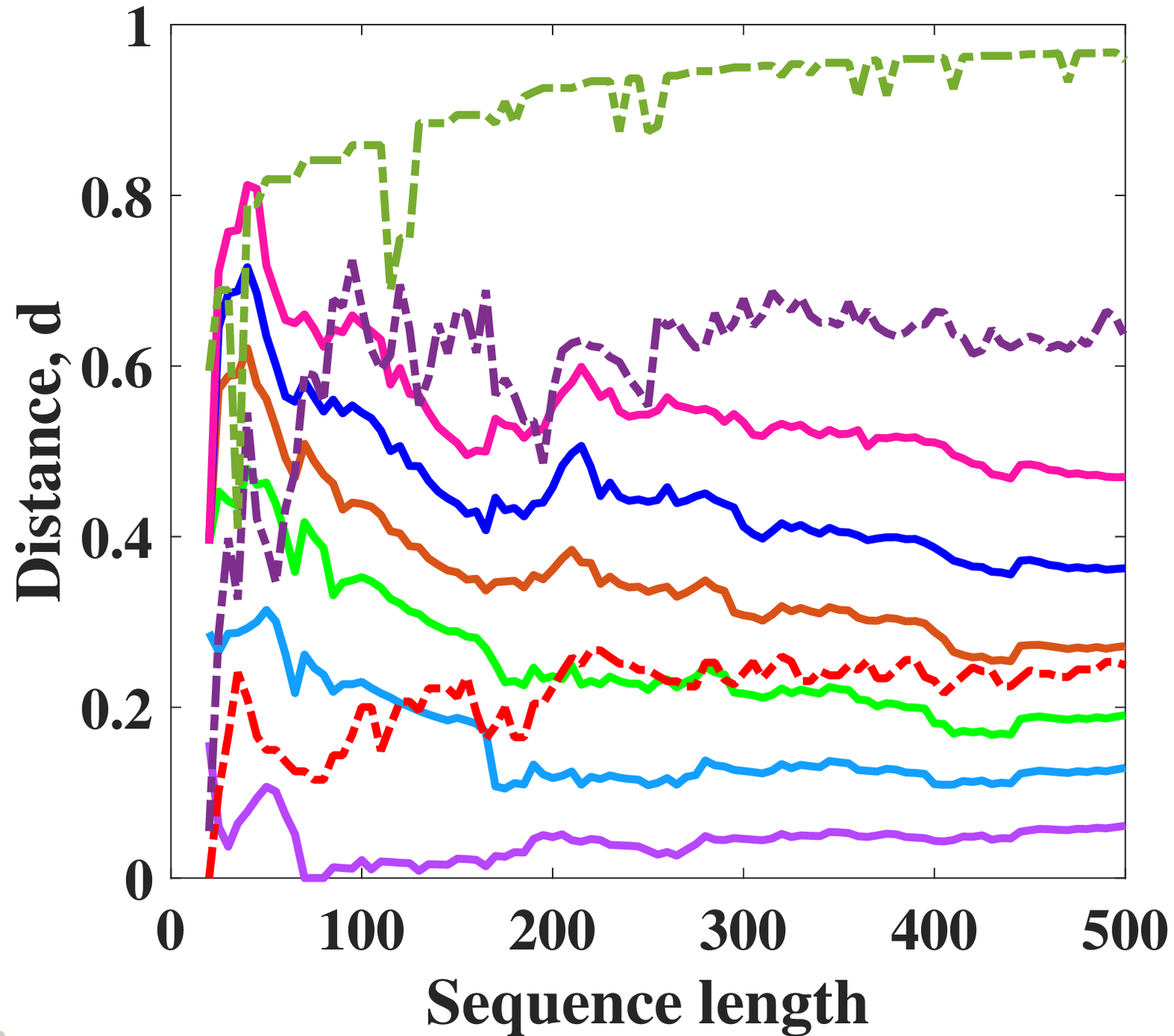}
\label{fig:nd3}\hspace{-1em}} 
\subfigure[]{
\includegraphics[width=4.2cm]{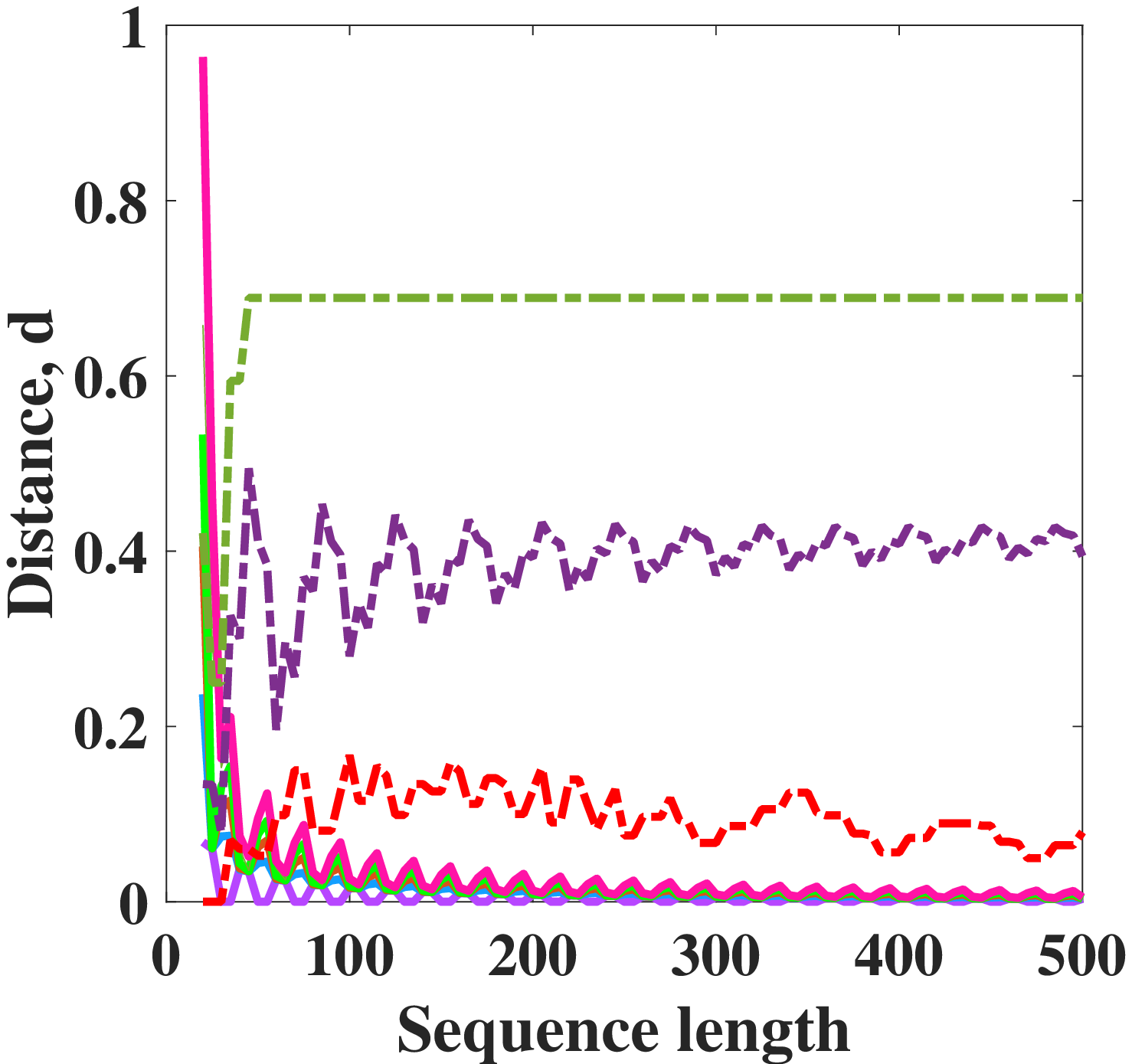}
\label{fig:nd4}}
\end{tabular} \vspace{-0.8em}
\caption{Temporal variation in distance based on LZ algorithms and empirical probability based methods for $k=0, 2, 4, 6, 8, 10$: (a),(b),(c),(d) for spike train pairs (i), (ii), (iii), (iv) shown in Figure \ref{fig:RPsynthetic} respectively.}
\label{fig:TempVariation}
\end{figure*}
\subsection{Proposed Empirical Dissimilarity Measure}

We computed the empirical contextual probabilities of symbols up to a pre-fixed maximum context length $k$ in each given spike sequence $X^n$ \cite{sathish2020}. For instance, consider the sequence $X^n=`010010111010100'$ and $k= 2$ (`010'- 4 - 4/5 : indicates symbol `0' occurs for four times when context `01' appears for five times in $X^n$):
\begin{itemize}
\item k=0 (no context):    `0' - 8 - 8/15 ; `1'- 7 - 7/15
\item k=1 :    `00' - 2 - 2/7 ; `01' - 5 - 5/7 ; `10' - 5 - 5/7 ; `11' - 2 - 2/7. 
\item k=2 :  `000 - 0 - 0; `001' - 1 - 1; `010' - 4 - 4/5 ; `011' - 1 - 1/5; `100' -  2 - 2/5 ; `101' - 3 - 3/5; `110' - 1 - 1/2; `111' - 1 - 1/2. 
\end{itemize}
The empirical Hellinger distance between probability distributions $P(x^n)$ and $Q(x^n)$ of two spike sequences is defined in terms of Bhattacharyya coefficient $BC(P,Q)$ as
\begin{equation}
d_{\mbox{\scriptsize EHD}} = \sqrt{1-\mbox{BC}(P,Q)}, 
\label{eq:dHD}
\end{equation}
where
\begin{equation}
\begin{aligned}
\mbox{BC}(P,Q) &= \sum_{x^n \in X^n} \sqrt{P(x^n)Q(x^n)} \\
 &=\sum_{x_1} \bigg[ 
 ...\bigg\{\sum_{x_{n-1}} \bigg(\sum_{x_n} \sqrt{P(x_n \vert x_1^{n-k-1}) Q(x_n \vert x_1^{n-k-1})}\bigg) \\ & \!\!\!\!\!\!\!\!\!\!\!\!
 \sqrt{P(x_{n-1} \vert x_1^{n-k-2}) Q(x_{n-1} \vert x_1^{n-k-2})}\bigg\} ...\bigg]\sqrt{P(x_1)Q(x_1)},
\end{aligned}
\label{eq:dBC} 
\end{equation}
and $P(x_n \vert x_1^{n-k-1})$ and $Q(x_n \vert x_1^{n-k-1})$ denote respective empirical conditional probabilities of the two sequences at context length $k$.

\subsection{Clustering using Gaussian mixture models}
\label{ssec:GMM}

We adopted the Gaussian mixture model for clustering the dissimilarities between the spike trains of different neuron pairs in population. In particular, the mixture probability density function (pdf) is assumed to follow 
\cite{mclachlan2000peel}
\begin{equation}
p(x;\theta) = \sum_{i=1}^{K} w_{i}        \phi(x;\mu_{i},\sigma^2_{i}),
\end{equation}
where $\phi(\cdot)$ indicates the Gaussian pdf, and $\mu_{i}$, $\sigma^2_{i}$ and $w_i$ respectively denote the mean, variance and mixing weight of the $i$-th Gaussian component. Further, the parameter vector
$\theta=\{\mu_{i}, \sigma^2_{i},w_i\}_{i=1}^K$, assuming $K$ Gaussian components.

\section{Results}
\label{sec:results}

\begin{figure*}[t!]
\centering
\includegraphics[width=\textwidth]{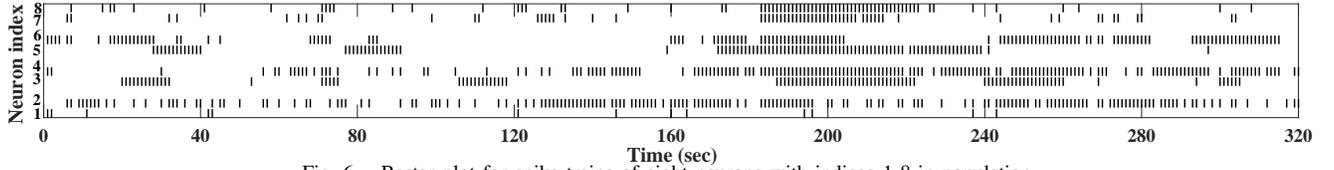}
 \vspace{-2.36em}
\caption{Raster plot for spike trains of eight neurons with indices 1-8 in population.}
\label{fig:RPneurons}
\end{figure*}
\begin{figure*}[t!]
\centering
\begin{tabular}{cc}
\includegraphics[width=0.9\textwidth]{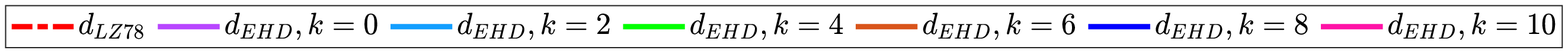}\\
\vspace{-1.8em}\\ 
\subfigure[]{
\includegraphics[width=4.2cm]{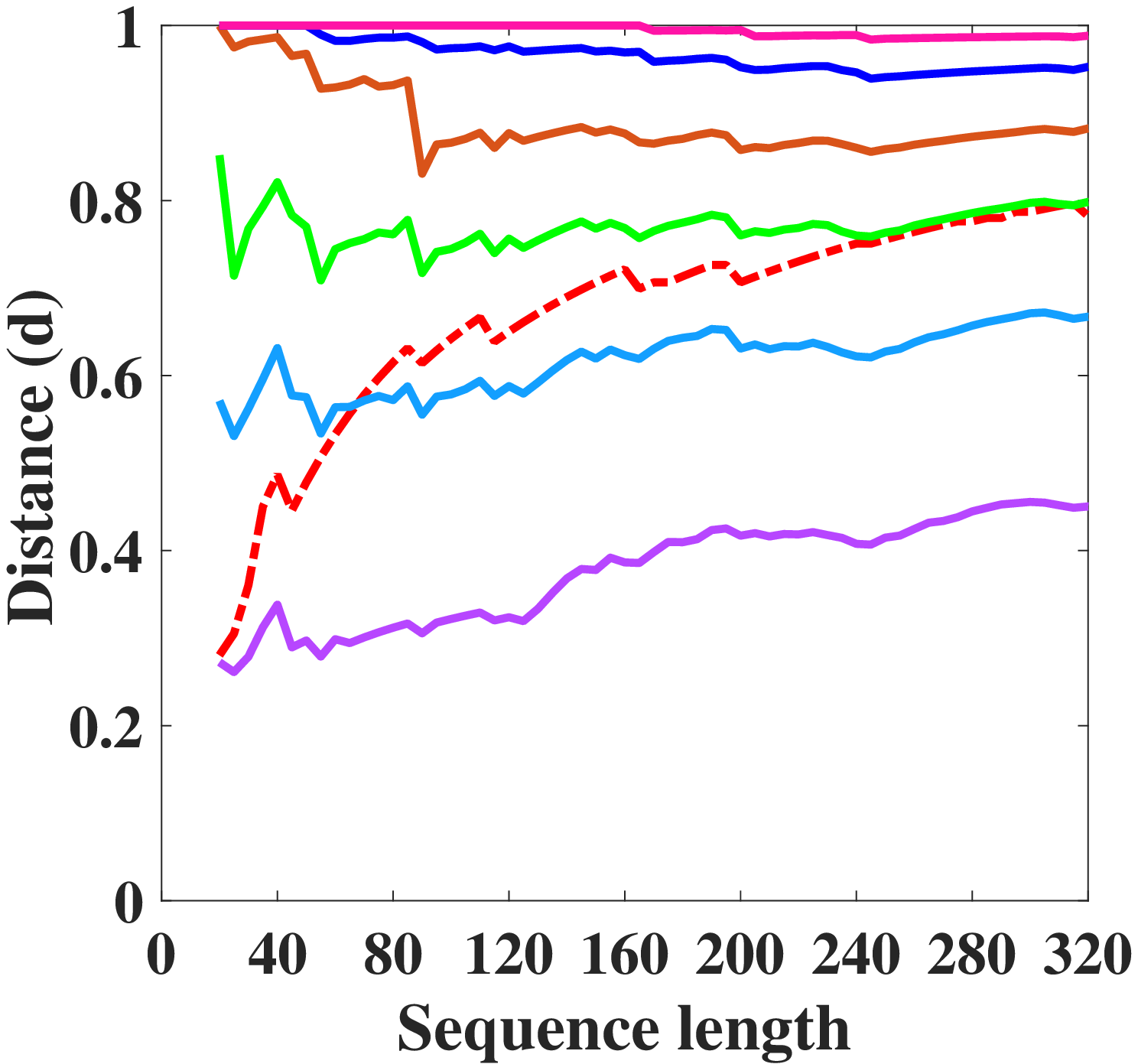}
\label{fig:12}\hspace{-1em}}
\subfigure[]{
\includegraphics[width=4.2cm]{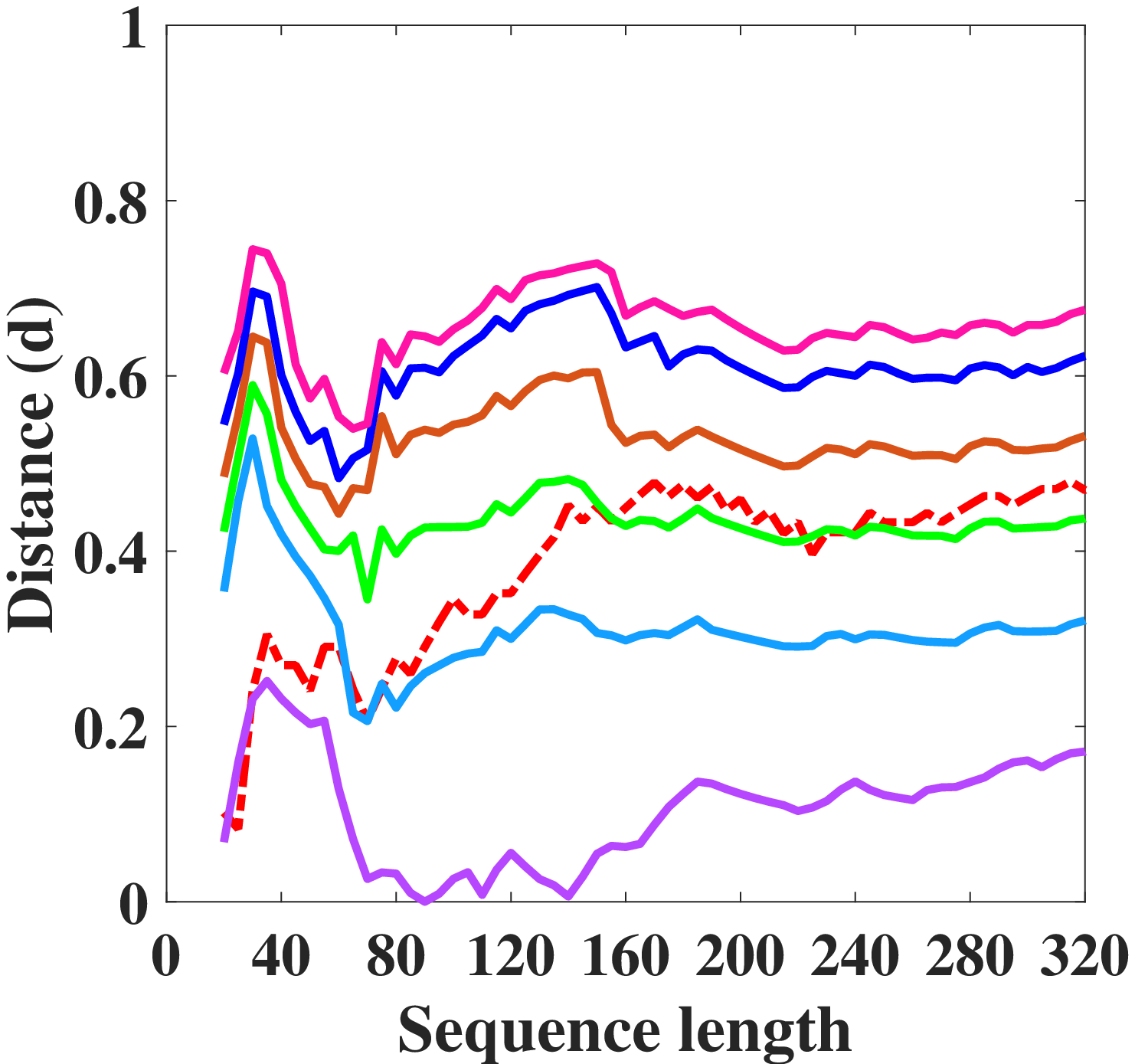}
\label{fig:34}\hspace{-1em}} 
\subfigure[]{
\includegraphics[width=4.2cm]{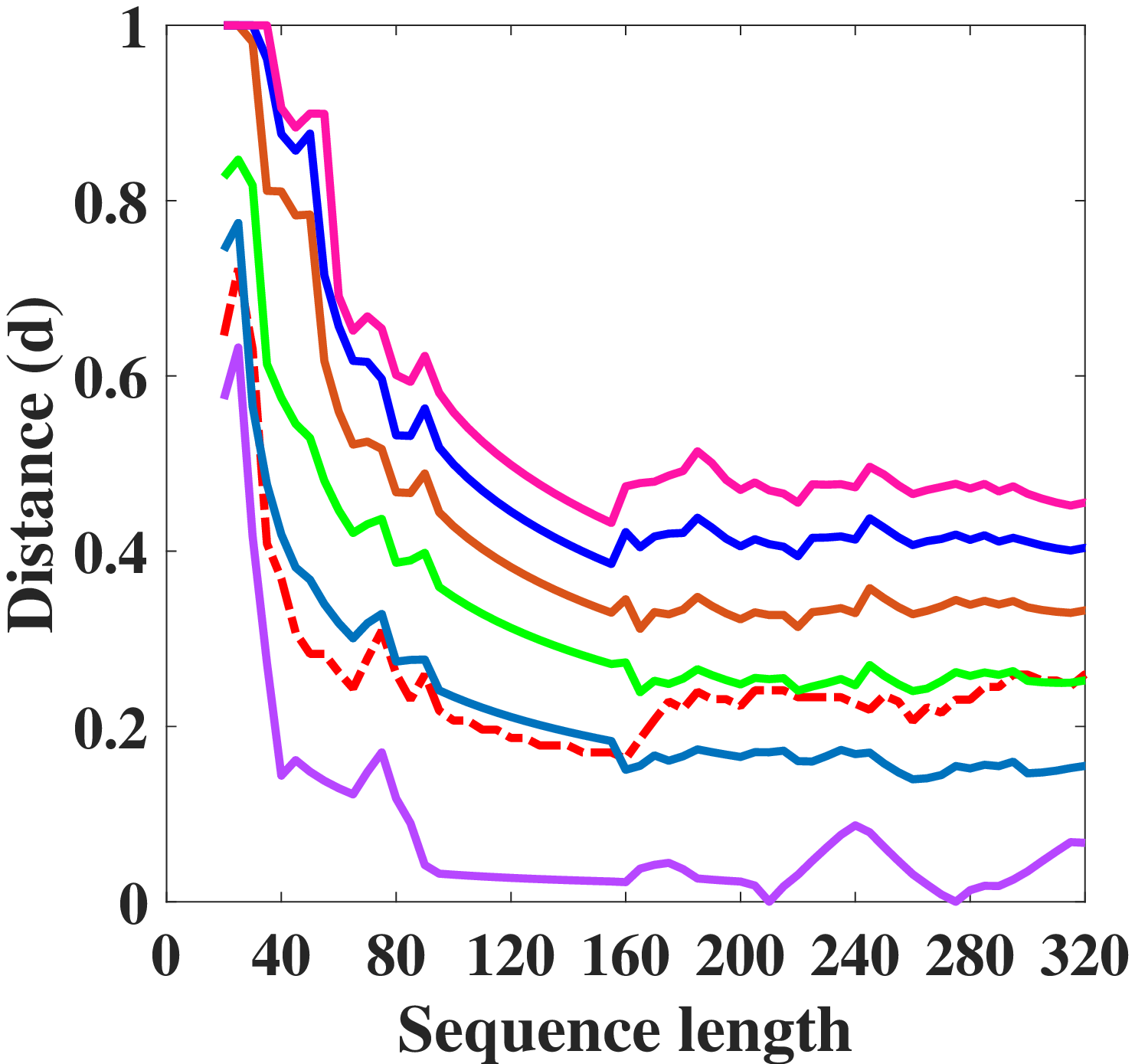}
\label{fig:56}\hspace{-1em}} 
\subfigure[]{
\includegraphics[width=4.2cm]{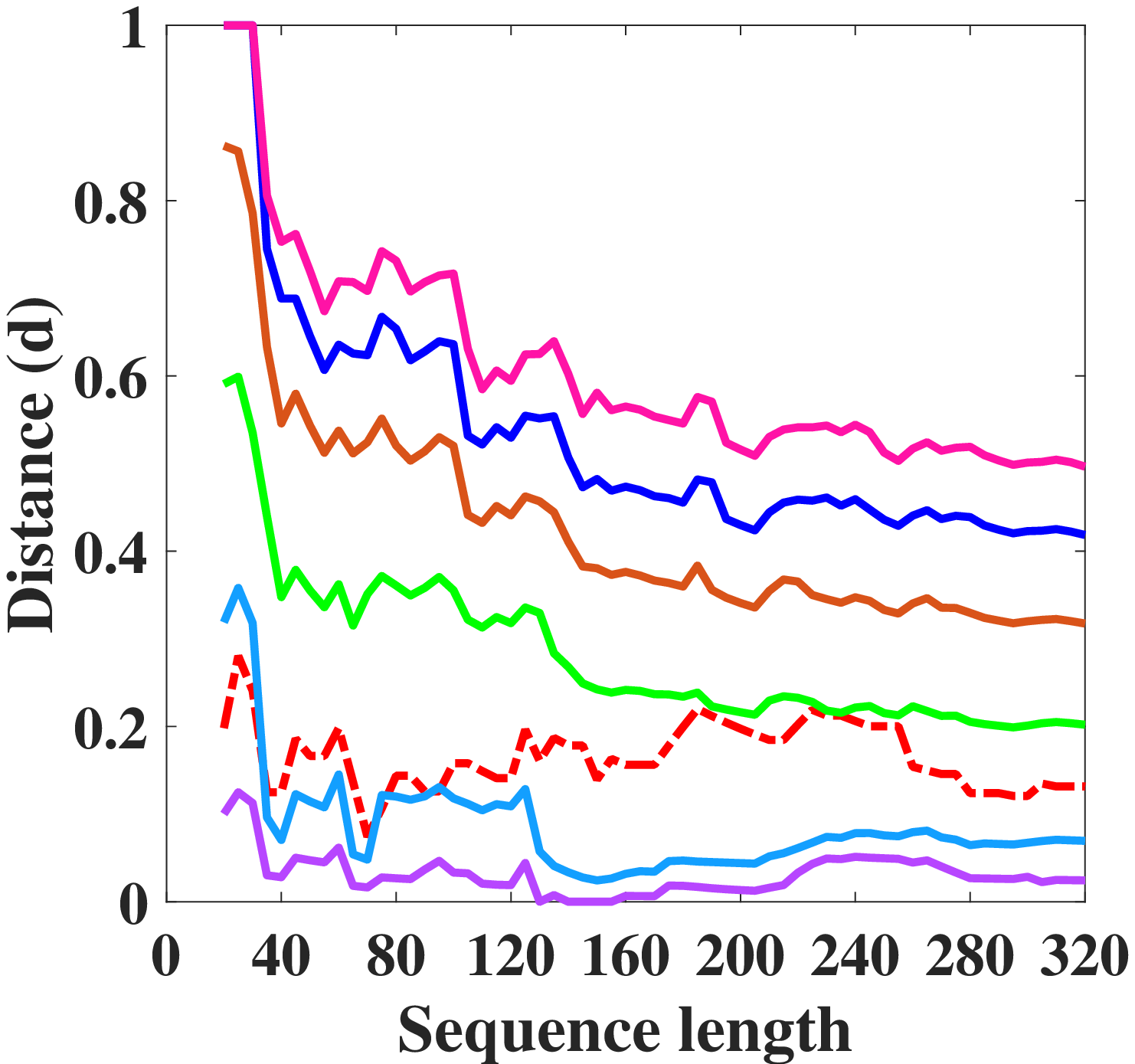}
\label{fig:78}}
\end{tabular} \vspace{-0.8em}
\caption{Temporal variation in distances based on LZ-78 algorithm and empirical probability based methods for $k=0,2,4,6,8,10$: (a),(b),(c),(d) for neuron pairs 1-2, 3-4, 5-6, 7-8, respectively}
\label{fig:N1to8}
\end{figure*}

We begin by demonstrating our tools on 
synthetic spike trains. Subsequently, such tools are applied to experimentally observed calcium spike trains.

\subsection{Synthetic Spike Trains}
\label{ssec:synthetic}

We generated four pairs of spike trains as follows:
(i) a pair following independently identically distributed ({\em iid}) Poisson model with firing rate $\lambda$= 50 $Hz$; (ii) an independent pair  
following Poisson model with firing rates 50 $Hz$ and 300 $Hz$; (iii) an {\em iid} pair, each following a Poisson model with a random $\lambda$ picked for each time window of length 50, and uniformly distributed in the interval [50  300] $Hz$; (iv) a phase-shifted pair of periodic spike trains with period 20 $ms$. In (i)-(iii), a spike in each $dt= 1ms$ interval is generated when a uniform random number in [0,1] turns out to be less than $\lambda dt$. The raster plot of all 8 spike trains, described above, is shown in Fig.\ref{fig:RPsynthetic}. We next compare empirical Hellinger distance measure $d_{\mbox{\scriptsize EHD}}$ for $k=0,2, 4, 6, 8, 10$ with existing LZ based distance measures $d_{\mbox{\scriptsize LZ78}}$, $d_{\mbox{\scriptsize LZ77}}$, $d_{\mbox{\scriptsize LZ76}}$ for (i)-(iv) as shown in Figs. \ref{fig:nd1}--\ref{fig:nd4} respectively. We computed  $d_{\mbox{\scriptsize EHD}}$ using (\ref{eq:dHD}) and $d_{\mbox{\scriptsize LZ78}}$, $d_{\mbox{\scriptsize LZ77}}$, $d_{\mbox{\scriptsize LZ76}}$ using (\ref{eq:dLZ}). 
In pair (i) (as well as in pair (iii)), the computed  distances $d_{\mbox{\scriptsize LZ78}}$ and $d_{\mbox{\scriptsize EHD}}$ are small as spike trains are {\em iid} (see Figs. \ref{fig:nd1} -\ref{fig:nd3}). However, the distances $d_{\mbox{\scriptsize LZ77}}$ and $d_{\mbox{\scriptsize LZ76}}$ are large. A similar behavior is observed in case of non-random periodic pair (iv) as well. So, $d_{\mbox{\scriptsize LZ77}}$ and $d_{\mbox{\scriptsize LZ76}}$ appear unsuitable as measures of distance, and will not be considered for further analysis.
Revisiting case (iv), $d_{\mbox{\scriptsize EHD}}$ quickly converges to zero, whereas $d_{\mbox{\scriptsize LZ78}}$ converges slowly. In case (ii), where rival spike trains have different distribution, 
$d_{\mbox{\scriptsize LZ78}}$ and $d_{\mbox{\scriptsize EHD}}$ both capture the dissimilarity. However, $d_{\mbox{\scriptsize EHD}}$ behaves similarly to $d_{\mbox{\scriptsize LZ78}}$ for suitable $k$ ($=6$), albeit with faster convergence.

\begin{figure}[t!]
\centering
\subfigure[]{
\includegraphics[width=\columnwidth, height=0.1\textheight]{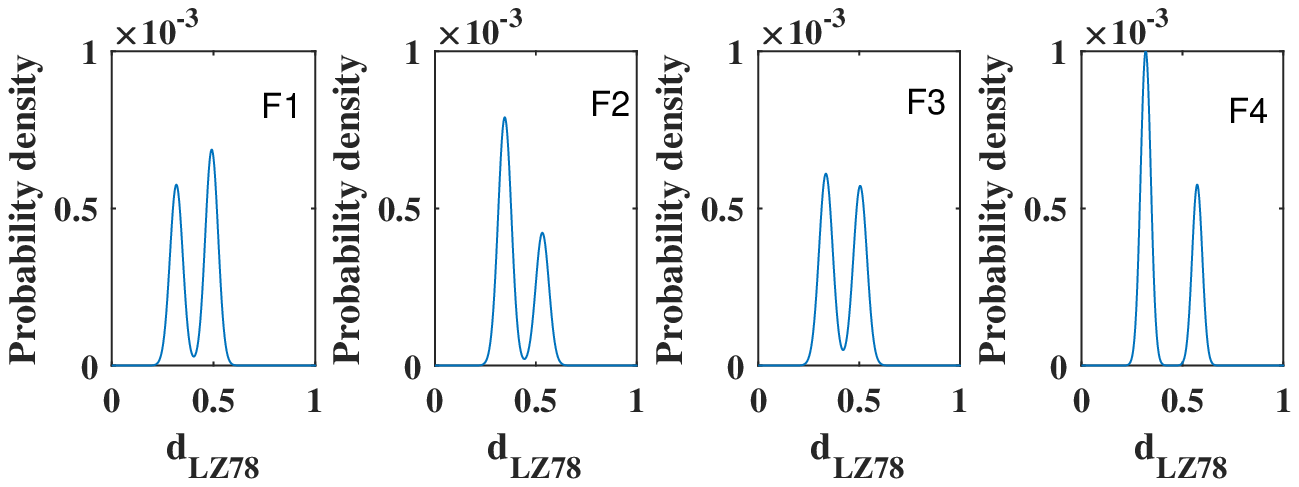}\label{fig:2dLZ78}} \vspace{-1em}\\
\subfigure[]{
\includegraphics[width=\columnwidth,height=0.1\textheight]{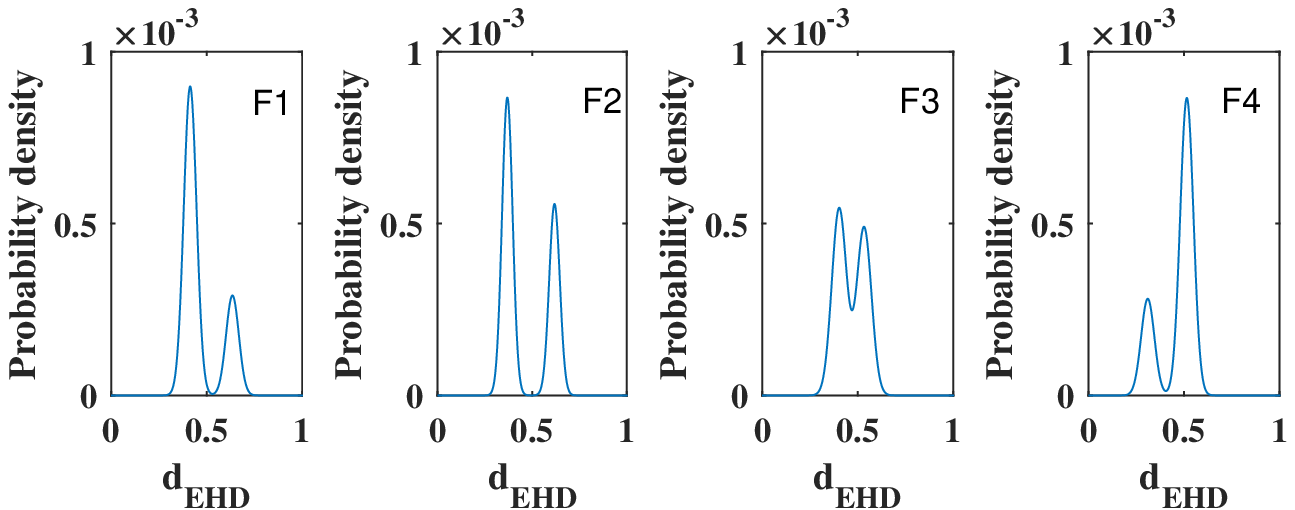}\label{fig:EHD}}
\vspace{-1.6em}
\caption{Variation in clustering with temporal shift: using  (a) $d_{\mbox{\scriptsize LZ78}}$ and (b) $d_{\mbox{\scriptsize EHD}}$ as  distance measure.}
\label{fig:2dclusters}
\end{figure}
\subsection{Calcium Spike Trains}
\label{ssec:experimental}

\begin{figure}[t!]
\centering
\hspace{-1.8em}
\subfigure[]{
\includegraphics[width=4.2cm]{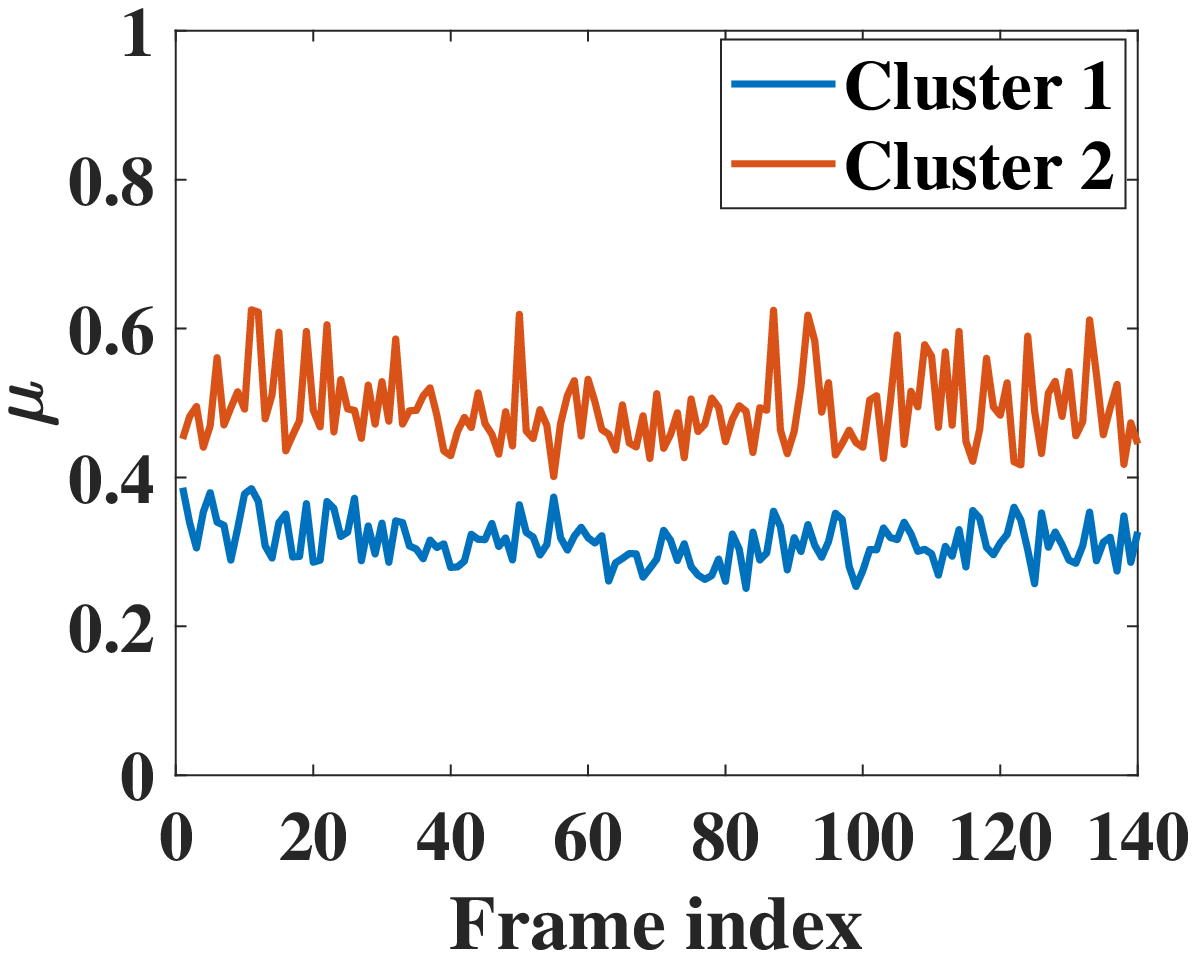}
\label{fig:mLZ78}\hspace{-1.5em}}
\subfigure[]{
\includegraphics[width=4.2cm]{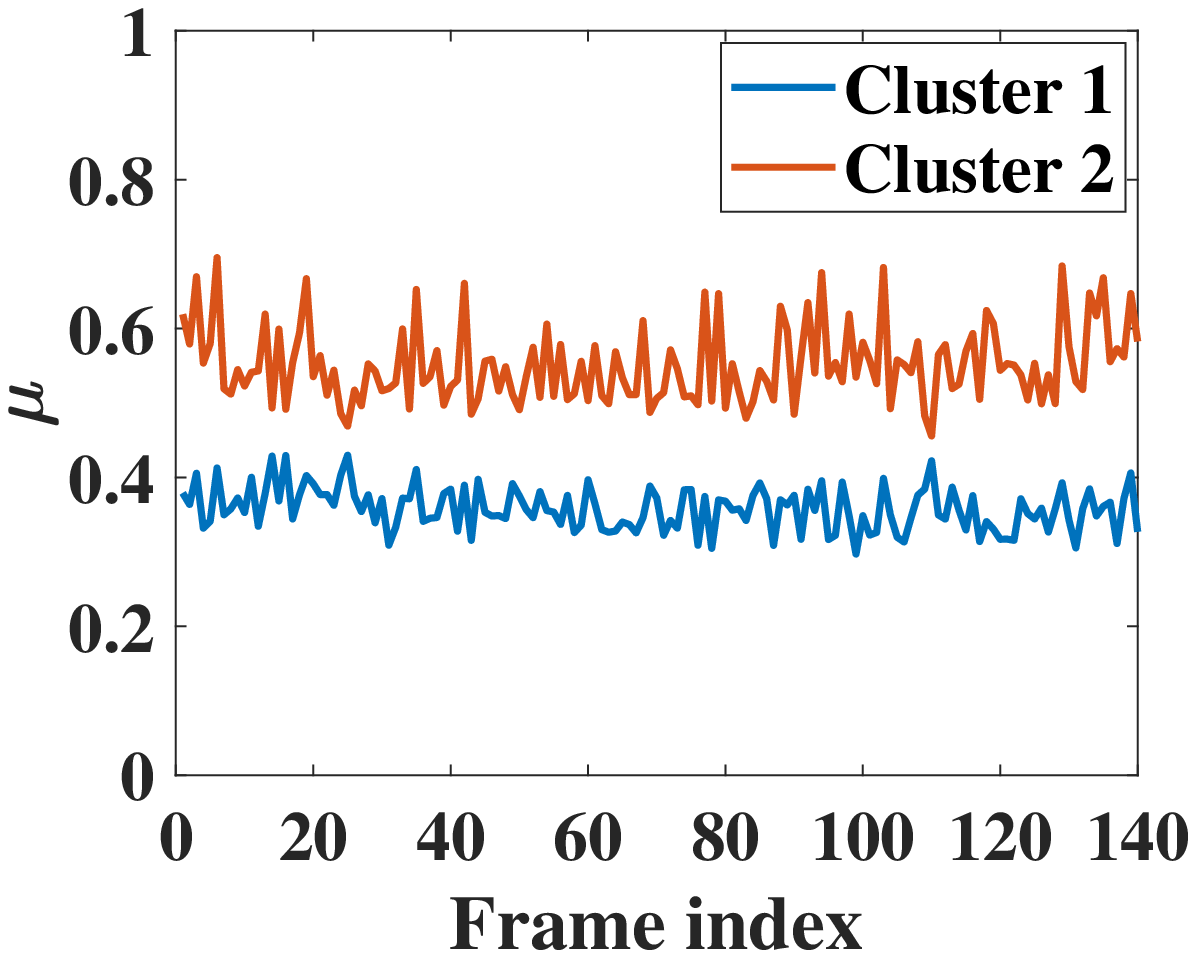}
\label{fig:mHD}\hspace{-1em}} \vspace{-1em}
\caption{Variation in cluster means ($\mu$) with temporal shift: using  (a) $d_{\mbox{\scriptsize LZ78}}$ and (b) $d_{\mbox{\scriptsize EHD}}$ as  distance measure.}
\label{fig:Cmean}
\end{figure}

\begin{figure}[t!]
\centering
\hspace{-1.8em}
\subfigure[]{
\includegraphics[width=4.2cm, height=0.1\textheight]{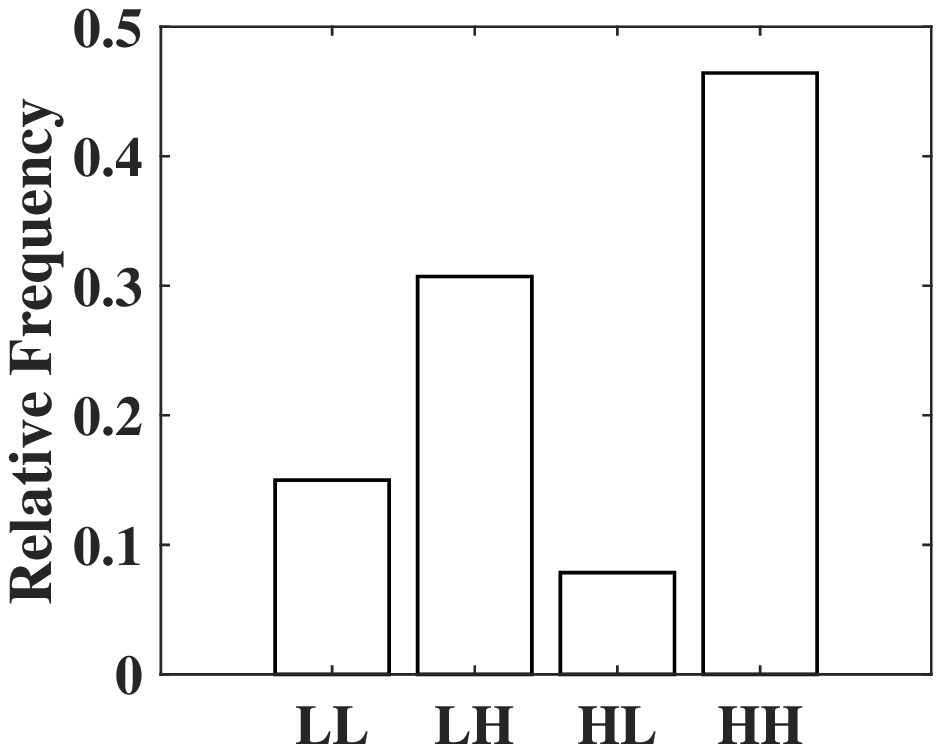}
\label{fig:pLZ78}\hspace{-1.5em}}
\subfigure[]{
\includegraphics[width=4.2cm,height=0.1\textheight]{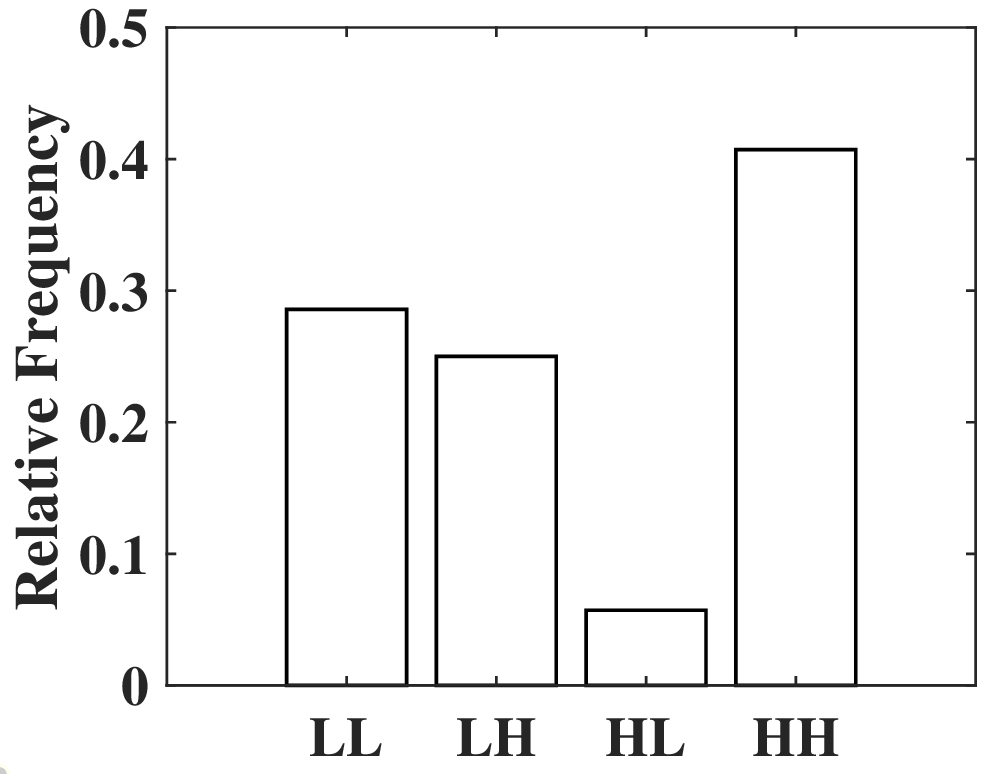}
\label{fig:pHD}\hspace{-1em}} \vspace{-1em}
\caption{Histogram of clusters LL, LH, HL and HH of cluster means under distortion measures (a) $d_{\mbox{\scriptsize LZ78}}$, and (b) $d_{\mbox{\scriptsize EHD}}$.}
\label{fig:pLH}
\end{figure}

As mentioned earlier, we consider 
eight neurons, indexed 1-8, whose spike trains are inferred. The corresponding raster plot is shown in Figure \ref{fig:RPneurons}. Taking successive neurons as pairs (1-2, 3-4, 5-6, 7-8), 
we next compared $d_{\mbox{\scriptsize EHD}}$ for $k= 0, 2, 4, 6, 8, 10$ with $d_{\mbox{\scriptsize LZ78}}$ for different sequence lengths in respective Figs. \ref{fig:12}-\ref{fig:78}. We observe that the proposed distance measure $d_{\mbox{\scriptsize EHD}}$ converges faster than $d_{\mbox{\scriptsize LZ78}}$ for all $k$. Further, $d_{\mbox{\scriptsize EHD}}$ approximates $d_{\mbox{\scriptsize LZ78}}$ for $k=4$ beyond sequence length 200. However, this analysis needs to be performed for more neurons to obtain optimum $k$. Next, we considered sequence length  200, and clustered neuronal spike train pairs using Gaussian mixture models based on distance measures $d_{\mbox{\scriptsize LZ78}}$ and $d_{\mbox{\scriptsize EHD}}$ with $k=4$. Varying number $K$ of clusters, we observed two significant clusters, which oscillated with temporal shift. Specifically, refer to Fig. \ref{fig:2dclusters} for such behaviour for both distance measures, when the starting frame is shifted by 0, 1, 2 and 3 samples. Interestingly, the variations in mean values of those clusters (for each of $d_{\mbox{\scriptsize LZ78}}$ and $d_{\mbox{\scriptsize EHD}}$), when plotted against temporal shifts ranging 0--140  (refer Figs.\ref{fig:mLZ78} and \ref{fig:mHD}, respectively), show oscillatory fluctuations that are small compared to the difference in mean. Further, we performed k-means clustering on means of the two clusters (labeled L, low-mean, and H, high-mean), each pair generated for temporal shifts 0--140, and similarly assigned a second label L or H. Next, we plotted relative histogram for classes LL, LH, HL and HH in Fig.\ref{fig:pLH} for both distance measures at hand. We observed that the relative count in LL in case of  $d_{\mbox{\scriptsize LZ78}}$ is less compared to $d_{\mbox{\scriptsize EHD}}$. It potentially indicates that  $d_{\mbox{\scriptsize LZ78}}$ shows some neuron pairs as statistically dissimilar, while those are actually similar.     




\section{Conclusion}
\label{sec:conclusion}
In this paper, we proposed an empirical Hellinger distance measure to quantify pairwise statistical dissimilarity in neuronal calcium spike trains. We illustrated our method with synthetic as well as experimental spike trains, and showed fast convergence. At the same time, our method exhibited similar (or, slightly superior) clustering behavior. The quantitative analysis of how fast is our method compared to existing methods would be considered as a future work. Further, we would like to extend our analysis for spike trains recorded with micro-electrode array \cite{mahmud2016processing}. We expect the proposed method to facilitate large-scale studies of functional clustering, especially involving short  sequences,  which  would  in  turn  identify  signatures  of various  diseases  in  terms  of  clustering patterns. In particular, the proposed scheme could assume importance in functional clustering of neurons in dissociated cultures.



\section{ACKNOWLEDGMENT}
We thank Drs. Mennerick and Gautam for providing materials and equipment. Sathish Ande thanks the Ministry of Electronics and Information Technology (MeitY), the Government of India, for fellowship grant under Visvesvaraya PhD Scheme.

\bibliographystyle{IEEEtran}
\bibliography{RefsBib}

\end{document}